\documentclass[twocolumn,pra,amssymb,showpacs]{revtex4-1}
\usepackage{graphicx}
\usepackage{amsmath}
\usepackage{subfigure}

\def\be {\begin{equation}}
\def\ee {\end{equation}}
\def\ba {\begin{eqnarray}}
\def\ea {\end{eqnarray}}

\def\bi {\begin{itemize}}
\def\ei {\end{itemize}}

\newcommand{\ben}{\begin{eqnarray}}
\newcommand{\een}{\end{eqnarray}}

\begin{document}

\title{Light Cones in Classical Dipole-Dipole Interacting Systems}

\author{Josep Batle$^1$}
\email{jbv276@uib.es}
\affiliation{
$^1$Departament de F\'{\i}sica, Universitat de les Illes Balears, 07122 Palma de Mallorca, Balearic Islands, Spain}

\author{Joan J. Cerd\`a$^2$}
\email{jj.cerda@uib.cat}
\affiliation{
$^2$Dpt. de F\'isica UIB i Institut d'Aplicacions Computacionals de Codi Comunitari (IAC3), Campus UIB, E-07122 Palma de Mallorca, Spain}

\author{Ph. Depondt$^3$}
\email{depondt@insp.jussieu.fr}
\affiliation{$^3$Institut des NanoSciences de Paris, UMR CNRS 7588, Universit\'e Pierre et Marie Curie-Paris 6, F-75252 Paris Cedex 05, France}

\author{J.-C. S. L\'evy$^4$}
\email{jean-claude.levy@paris7.jussieu.fr}
\affiliation{$^4$Laboratoire Mat\'eriaux et Ph\'enom\`enes Quantiques, Universit\'e de Paris CNRS, F-75013, Paris, France}

\date{\today}

\begin{abstract}

The speed at which the magnetic interaction propagates along a chain of classic dipoles is discussed here. While in 
the quantum information counterpart for long-range interacting spins, where the speed of propagation of the information plays 
a paramount role, it is not strictly clear whether a light cone exists or not, here we provide numerical evidence that 
interacting dipoles do posses a linear light cone shortly after a perturbation takes place. Specifically, a power-law expansion 
occurs which is followed by a linear propagation of the associated interaction. As opposed to the quantum case, and 
in analogy with the so-called {\it speed of gravity problem}, we find that the speed of propagation of information can be 
arbitrarily large in the classic context. In order to agree with special relativity, we propose the derivation of a frame-independent 
Landau-Lifshitz equation.

\end{abstract}

\maketitle

The speed at which one can process quantum information by means of unitary operations is constrained by
the Lieb-Robinson bound in all its possible forms \cite{liebrobinson1972}, and depends on the nature of the interactions in systems of 
(short or long-ranged) spins. Arrays of spins act as buffers or bus channels for, among other tasks, 
transmitting quantum information.

In a non-relativistic setting, after a time $t$ once a perturbation occurs, a local chance appears to be extremely small outside 
an effective light cone, at least for systems forming lattices whose constituents interact with finite-range forces. This situation is known as 
{\it quasi-locality} and is the main subject of the aforementioned Lieb-Robinson bound.

In the case of short-range interacting spin systems, the information propagation is 
restricted inside a ``linear light cone.'' However, when considering long-range counterparts, linear light cones become rather subtle, for 
long-range interactions, per definition, allow instant communications between distant parties. 
By long-range interaction one understands that the interaction strength between separated sites decays as $L^{-\alpha}$ with the distance $L$. 
Depending on the exponent $\alpha$, both the linear and nonlinear light cones can appear. 
One clear outcome of the previous situation is the so-called {\it linear light cone problem}, which 
clarifies whether linear right cones can exist in long-range interacting systems, and what is the general criterion for it. Arguably, it is one 
of the most important and intriguing open problems in quantum information processing.

Among all Lieb-Robinson bounds, Foss--Feig et al. \cite{FossFeig} proved that the effective light cone is at most polynomial with 
respect to time. Today, to the best of our knowledge, no critical value of $\alpha$ to obtain a linear light cone has been found. 

The study of the Lieb-Robinson theorem via its bond has been, and currently is, the subject of intense research \cite{kastoryano2013,descamps2013,gong2014,maghrebi2016,iyoda2017,chanda2018,elgart2018,tran2020,nachtergaele2006,bravyi2006,
nachtergaele2009,eisert2009,raz2009,premont2010}. 
Surprising enough, no one has focused the problem from the classical perspective. Very few instances admit in the classic realm 
the study of the propagation of information along a medium besides contact forces. One such case is the so-called {\it speed of gravity}, which 
can only be resolved within the framework of general relativity \cite{rham2020}. Usually, the quantum case explores systems 
that interact \`a la Heisenberg, with all the coupling constants following a power-law behavior, and conclusions are drawn from the exponent $\alpha$ of the 
interaction and the dimension $D$ where the system is embedded \cite{gong2014,tran2020,elgart2018,chanda2018,maghrebi2016}.

Luckily, there is a physically-motivated classical counterpart to the problem defined 
quantum-mechanically, namely, that of an array of dipoles solely interacting via the dipole-dipole force. 
Properties of magnetic materials formed by dipoles depend crucially on the nature of the dipole-dipole interaction. In magnetically ordered crystals, 
the ground state of the spin system is determined by the exchange interaction. There, the role of dipole-dipole force is, among other effects, to 
stimulate the formation of domains.
Now, at this juncture, we have a ``natural'' interaction decreasing 
with $\alpha=3$, whose leading term resembles the Heisenberg one. Therefore, the study of the dipole dynamics in an array of dipoles will become the main goal of the present 
work. Fortunately, there exists an equation, derived by Landau and Lifshitz \cite{landau1992}, that can describe the time evolution of each dipole. However, as opposed to the quantum case, 
one needs to stress the fact that the spatial dimension, regardless of the arrangement of dipoles, is {\it always} $D=3$. This will have consequences for 
no Lieb-Robinson bound could be applied in this case (conditions relating $\alpha$ and $D$ need not hold in our case).

{\it Statement of the problem}.-- The Landau-Lifshitz equation that governs the dipole dynamics is given by

\begin{eqnarray} \label{LL}
\frac{d\overrightarrow{s_l}}{dt} \,&=&\,- \overrightarrow{s_l} \times\, \overrightarrow{H_l}, \cr
\overrightarrow{H_l}\,&=&\,C_M\,\sum \limits_{l'\ne l} 
\bigg( 3\frac{\overrightarrow{s_{l'}}\cdot \overrightarrow{r_{l,l'}}}{||\overrightarrow{r_{l,l'}} ||^2} \overrightarrow{r_{l,l'}} \, -\, \overrightarrow{s_{l'}} \bigg)
\,\frac{1}{||\overrightarrow{r_{l,l'}} ||^3}, \cr
\frac{1}{C_M} \, \frac{d\overrightarrow{s_l}}{dt} \,&=&\, \mathcal{A} \cdot \overrightarrow{s_l}, 
\end{eqnarray}

\noindent where the evolution for individual dipoles is explicitly shown (third equation), $C_M=\frac{\mu_0}{4\pi}$, and the corresponding matrix $\mathcal{A}$ is given by

\begin{align}
 \begin{pmatrix}
 0 & \sum \limits_{l'\ne l} \frac{1}{||\overrightarrow{r_{l,l'}} ||^3} s_{l'}^{(z)}  & 2\sum \limits_{l'\ne l} \frac{1}{||\overrightarrow{r_{l,l'}} ||^3} s_{l'}^{(y)} \\
 - \sum \limits_{l'\ne l} \frac{1}{||\overrightarrow{r_{l,l'}} ||^3} s_{l'}^{(z)} & 0 & \sum \limits_{l'\ne l} \frac{1}{||\overrightarrow{r_{l,l'}} ||^3} s_{l'}^{(x)} \\
 - 2\sum \limits_{l'\ne l} \frac{1}{||\overrightarrow{r_{l,l'}} ||^3} s_{l'}^{(y)} & - \sum \limits_{l'\ne l} \frac{1}{||\overrightarrow{r_{l,l'}} ||^3} s_{l'}^{(x)} & 0 \\
 \end{pmatrix}.
\end{align}

\noindent In a one dimensional system parallel to the $x$ axis, $\mathbf{e}_{l,l'} \parallel \mathbf{e}_x$ and
\begin{equation}
\mathbf{H}_l = \sum_{l' \neq l} \frac{\mathbf{s}_{l'} -3s_{x_{l'}}\mathbf{e}_x} {|r_{x_{l,l'}}|^3}
\end{equation}
One way to see that the magnitude of each dipole is preserved is via the nature of $\mathcal{A}$. This matrix is a skew-symmetric one, which implies that 
all non-zero eigenvalues come in pairs or purely imaginary numbers and, in turn, guarantees that $|| \overrightarrow{s_l} (t)||$ remains constant $\forall \,\,l,t$. Since no 
damping term is considered in (\ref{LL}), the total dipole-dipole energy is preserved. 

Let us, for the sake of numerical ease, normalize to unity $\overrightarrow{s_l}(t)=(s_{l}^{(x)}(t),s_{l}^{(y)}(t),s_{l}^{(z)}(t)) $ and take distances in terms of $a$, the interdipole distance. Then $C_M$ 
becomes $\frac{\mu_0}{4\pi}\frac{m}{a^3}$, where $m$ is the magnitude of the dipole and $a$ the interdipole distance. Incidentally, and without loss of generality, 
$m$ can be numerically adjusted such that $C_M$ is equal to one.

\begin{figure}[htbp]
\begin{center}
\includegraphics[width=8.8cm]{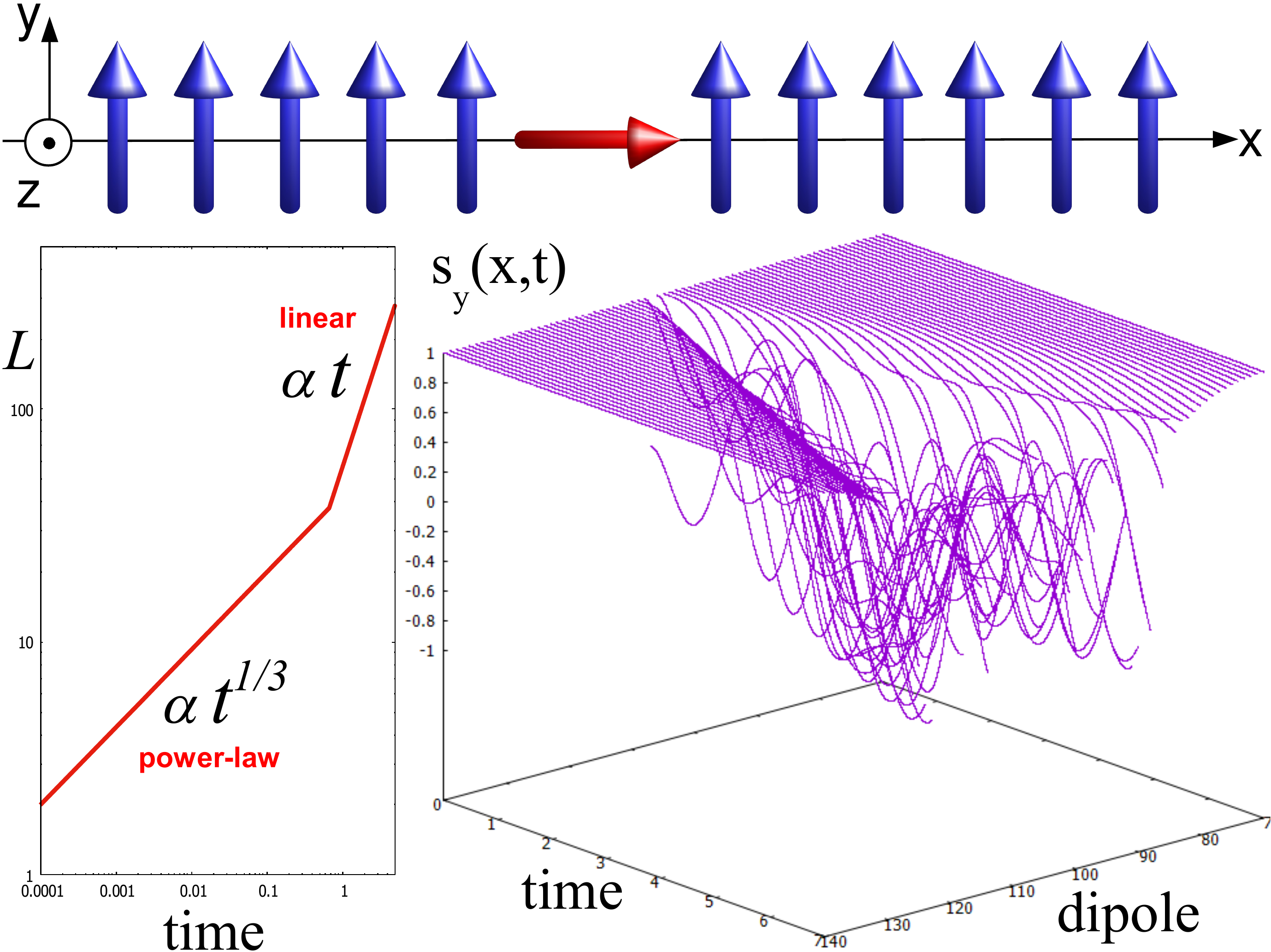}
\caption{(Color online) Initial configuration for $N=213$ dipoles. The dipole at position $x\,=\,106$ is perpendicular to the rest. Bottom left: the distance travelled 
by the interaction from dipole $l=106$. For initial times, they follow a 
power law evolution, which later becomes linear with time. Specifically, we have i) $x(t)\,=\,106\,\pm\,43\,t^{1/3}$, followed by ii) $121\,\pm\,56\,t$. Bottom right: the free evolution for 
$s_{l}^{(y)}(x,t)$ clearly shows the 
typical V-shape of a linear light cone. Positions are given in units of $a$. See text for details.}
\label{fig1}
\end{center}
\end{figure}

We shall consider two important configurations for the dipoles. The first one consists of an finite chain of $N$ dipoles ($N$ sufficiently big enough) along the $x$ direction, as described in Fig. 1. Suppose that the total number of dipoles is odd $N=2M+1$. We do not choose a setting where a perturbation starts 
at the beginning of the chain in order to avoid any finite-size effects. Thus, to such an end, at $t=0$ we will set all dipoles pointing upwards, and the one in the {\it middle}, that is, at position $l=M$, pointing 
perpendicularly to the rest. In this fashion, we will reproduce all results as if they occurred inside the bulk, and therefore avoiding the ends provided $N$ is large enough. 
Specifically, $\overrightarrow{s_l}(t=0)=(0,1,0)\,\,\forall \,\,l \ne M$ and $\overrightarrow{s_{l=M}}(t=0)=(1,0,0)$. Notice that no time-dependency would occur classically in (\ref{LL}) if there was no projection of dipole $l=M$ along the $y$ axis, as opposed to the quantum case. Needless to say, all results will be symmetric at both sides of $l=M$. 
The second instance considered is given in Fig. 2, with $\overrightarrow{s_l}(t=0)=(1,0,0)\,\,\forall \,\,l \ne M$ and $\overrightarrow{s_{l=M}}(t=0)=(0,1,0)$.
At this point, one has to bare in mind that the speed of the dipole-dipole interaction is infinite classically, as seen in (\ref{LL}). However, when dipoles start to move, their interaction mediated among themselves is not instantaneous.

At this juncture, and previous to any numerical computation, we shall clarify how the detection of information propagation translates in practice. In quantum information, the fidelity quantity 
$\mathcal{F}=||\langle \Phi|\Psi\rangle||$ measures how similar two quantum states are. Classically, and speaking of dipoles, in the first case of Fig. 1, we need to assess the first dipole $l^*$ such that at the time 
$t=t'$, the classical fidelity \cite{uhlmann2011} $\mathcal{F} \equiv (s_{l^*}^{(x)}(t'),s_{l^*}^{(y)}(t'),s_{l^*}^{(z)}(t')) \cdot (0,1,0)=s_{l^*}^{(y)}(t')$ starts to differ from 1. Let us define $\mathcal{F} \equiv 1\,-\,\delta$. Thus, 
the best that can be done is to compute the speed of propagation of information $v_s$ along the chain of interacting dipoles from $\Delta L = l^* - M$ and $t=t'$ {\it as soon as} $s_{l^*}^{(y)}(t') \leq 1\,-\,\delta$. This situation has profound consequences in defining an ``exact" speed of propagation $v_s$, for it depends on the value of $\delta$. In point of fact, and for any numerical scheme, the resolution of Eq. (\ref{LL}) 
will essentially depend on the time step $ht$ regardless of the desired accuracy. In other words, $s_{l^*}^{(y)}(t') = 1\,-\,\delta$ will always depend on $ht$ via $\delta=f(ht)$ in an extremely non-trivial fashion. This 
means to directly ``detect" the interaction heavily depends on the machine precision, and it is doomed to fail in order to provide a unique answer. However, although not shown here, this {\it direct} method qualitatively describes 
what is found later on.
In order to obtain quantitative results, what shall be done in practice is to adjust on the position-time plane of the evolution of the dipoles a certain curve. The quantity over which we will consider constant values will be 
that of $1\,-\,\mathcal{F}$. Obviously, the lowest value provided by the numerical computations will be template over which we shall infer the propagation of the information.

{\it Results}.-- The details of the numerical resolution of the time evolution of all dipoles in all settings is described in the Supplementary Material. What is relevant is that the numerical scheme guarantees, 
both the preservation of the total energy of the system and the magnitude of all time-evolved dipoles. Additionally, $C_M$ is set to 1. 

The first instance that we shall consider is the configuration of $N=213$ dipoles as shown in Fig. 1. The middle dipole points towards the positive $x$ axis, whereas all the rest point upwards. Although 
it is not maximal, this configuration is of high energy per dipole. The numerical solution of the coupled set of equations (\ref{LL}) returns that the position travelled from the central dipole by the interaction evolves as $x(t)\,=\,A\cdot t^{\beta}\,\cdot a$ ($a$ is the interdipole distance) initially, which is followed by a linear expression of the type $B\,+\,v_s\,t$, where $v_s$ is the speed of propagation of the information.  
$v_s$ is, generally, of $O(10)$ in units of $a$ per second. To be more precise, we obtained $x(t)\,=\,106\,\pm\,43\,t^{1/3}$, followed by $121\,\pm\,56\,t$

The speed $v_s(t)\,=\,\frac{d\,x(t)}{dt}\,=\,A\cdot\beta\cdot t^{\beta\,-\,1}\,a$ is initially infinite, and eventually becomes constant, usually beyond $t\,=\,0.1$. The speed $v_s$ at $t=ht$ is 
$A\cdot\beta\cdot ht^{\beta\,-\,1}\,a$, which can be arbitrarily large as compared to the speed of light, which is in flagrant contradiction with the tenets of special relativity.  

We have deliberately avoided finite-size effects in the system by not considering the interaction when it reaches the last dipoles at both ends. Considering larger chains only extends the linear regime. Therefore, our results can be regarded as effectively occurring in the bulk. Also, some further analysis shows that $v_s$ in the linear light cone is indeed proportional to $m$, the magnitude of the dipole. Obviously, 
$m=0$ would imply considering no dipoles and, hence, no possible propagation. The bare analysis of the fidelity in Fig. 1 clearly shows the typical V shape of the linear light cone, as it extends towards 
the ends of the dipole lattice.

Another setting shall be given as follows: in a chain of $N=2M+1$ dipoles, $\overrightarrow{s_l}(t=0)=(0,1,0)\,\,\forall \,\,l \ne M$ and $\overrightarrow{s_{l=M}}(t=0)=(0,0,1)$, as depicted in Fig. 2 for $N=257$ 
dipoles. Notice that this configuration corresponds to the minimum energy of the systems, when all dipoles are aligned. Proceeding as before, we find $x(t)\,=\,129\,\pm\,40\,t^{1/3}$ followed by 
$141.5\,\pm\,57\,t$. Notice that not only the functional form coincides with the case in Fig. 1, but the coefficient $A$ for $t^{1/3}$ and the speed $v_s$ are almost the same in both cases, respectively. 
Also, if we assume that the propagation of the interaction occurs continuously, imposing continuity and derivability in both regimes we obtained that the term $B$ in the linear regimes goes as 
$B\,=\,\frac{2}{\sqrt{27}}\,\frac{1}{\sqrt{v_s}}\,A^{3/2}$. Thus, the only two quantities needed to quantitatively describe the propagation of the information along the chain, in both two regimes, are 
$A$ and $v_s$.

It is remarkable that the speed of the linear light cone $v_s$ seems to not depend on the particular dipole configuration. In the setting of Fig. 1, the overall dipole-dipole energy interaction is greater than the one in Fig. 2, which 
is in fact the ground state in the thermodynamic limit. Thus, $v_s$ seems to be a quantity that solely depends on the dipole-dipole interaction nature, as well as the $A$ accompanying $t^{1/3}$.

The evolution of the fidelity $\mathcal{F}$ (or $1\,-\,\mathcal{F}$) for all dipoles, is shown in Fig. 2. Interestingly enough, this configuration leads to two solitons moving away from the central dipole at further 
times. When the interaction reaches the ends of the chain, the dipoles start to bounce back, precisely at the same speed at which the solitons are moving. 

Admittedly, the previous method for finding the functional form for the time evolution may appear somehow artificial, although it is numerically correct. These result does not really come as a surprise: our initial configuration (in the second case) is with all dipoles parallel to $x$ while the source, which for convenience we can place at the origin, is parallel to $z$. Therefore the field felt by a dipole at location $x$ writes, with all relevant constants set to one:
$$\mathbf{H}(x) = \frac{\mathbf{s}(0)}{x^3}= \frac{s_z(0)}{x^3}\mathbf{e}_z$$
Now, this should be inserted into the Landau-Lifshitz equation (\ref{LL}).
At the onset of the motion, we can consider $s_z(0)=1$ to be constant as it evolves as a cosine : $1-\gamma^2/2$ where $\gamma$ is the angle it makes with the $z$ axis. Thus:
$$\frac{d\mathbf{s}(x,t)}{dt} = -  \frac{s_z(0)}{x^3}\mathbf{e}_z\times \mathbf{s}(x,t)$$
and:
$$\frac{ds_y(x,t)}{dt} = -\frac{1}{x^3} s_x(x,t)$$
again, $s_x(x,t)$ evolves as the cosine of the angle with the $x$ axis, and can be considered constant $=1$ for short time intervals $dt$, finally
$$x = \left( \frac{dt}{ds_y}\right)^{\frac{1}{3}}$$
That is, the distance $x$ from the source at which we observe a change $ds$ increases as $dt^{\frac{1}{3}}$.

The same reasoning with $H\propto x^{-\alpha}$ would yield
\begin{equation}
x = \left( \frac{dt}{ds_y}\right)^{\frac{1}{\alpha}}\label{precursor_eq}
\end{equation}
and it so turns out that a simulation with $H \propto 1/r^4$ does indeed produce a $dt^{\frac{1}{4}}$ behavior. Equation (\ref{precursor_eq}) thus 
appears to be general for very short time intervals. This is certainly of paramount importance when compared to the quantum case: shortly after propagating, 
the information will travel along the classic chain of dipoles as $A\,t^{\frac{1}{\alpha}}$, for interactions decaying as $1/r^{\alpha}$.

The dependence of $v_s$ on the magnitude of the dipole is apparent from closer inspection of the Landau-Lifshitz equation (\ref{LL}). The transformation $t\,\longrightarrow\,\frac{\mu}{4\pi}\,\frac{m}{a^3}\,t$ 
implies that $v_s\,\propto\,\frac{\mu}{4\pi}\,\frac{m}{a^3}$. Thus, the greater the magnitude of the dipoles, the faster the propagation of the interaction. Evidently, it also applies to changing the magnetic permeability 
and/or the interparticle distance. Remarkably, the value of $B\,=\,\frac{2}{\sqrt{27}}\,\frac{1}{\sqrt{v_s}}\,A^{3/2}$ remains invariant under such rescaling.

We need to conciliate the finite speed of the magnetic interaction with the bound of information propagation defined by the speed of light $c$. This can be introduced $ad-hoc$ 
in the form of a ``retarded" interaction, with $t\,\rightarrow\,t\,\pm\,\frac{1}{c}\, ||\vec{r}-\vec{r}_{source}|| $ plus a covariant formulation for the Landau-Lifshitz equation.

\begin{figure}[htbp]
\begin{center}
\includegraphics[width=8.8cm]{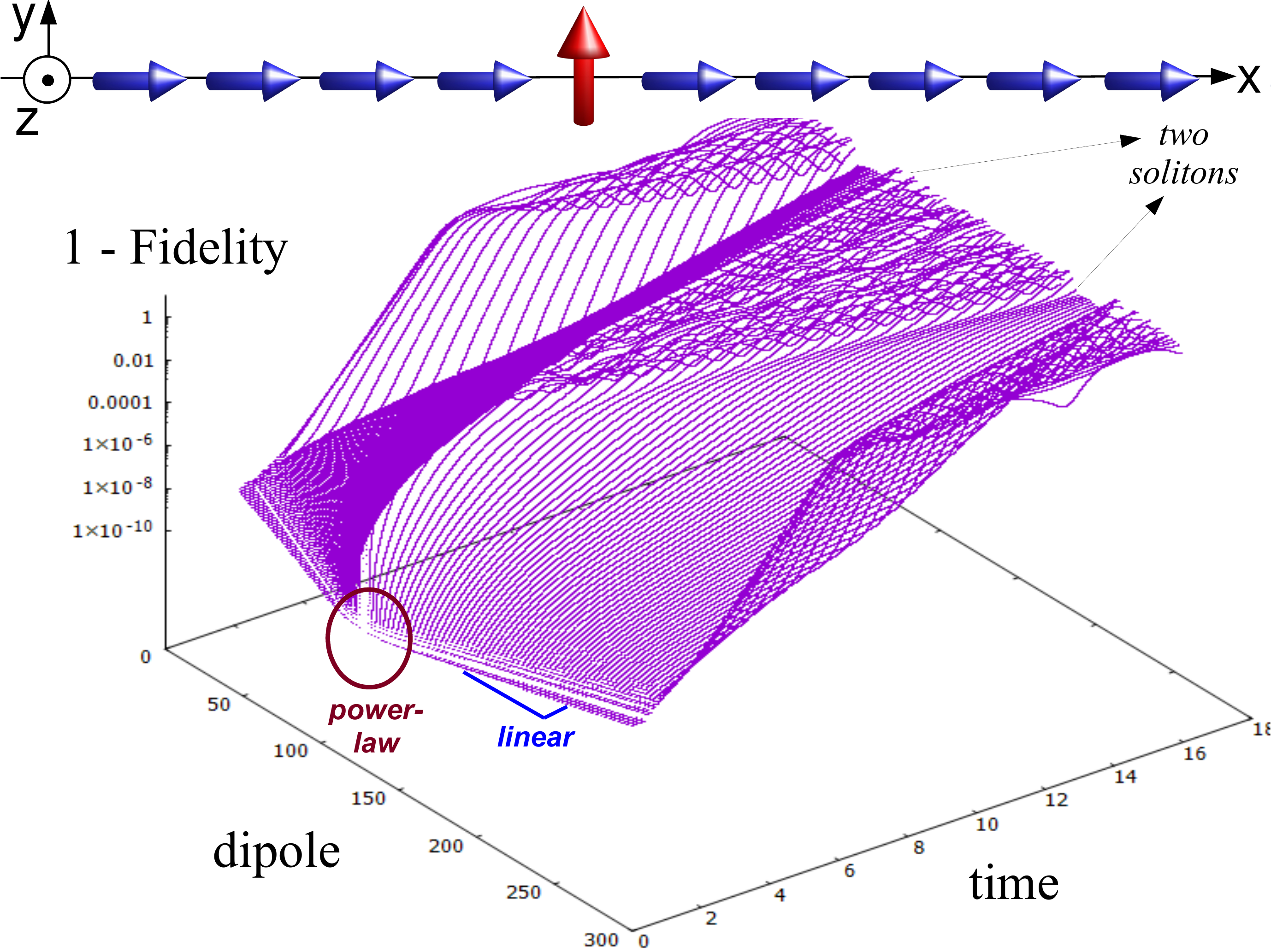}
\caption{(Color online) Similar plot to Fig. 1 but for another initial configuration for $N=257$ dipoles. In the free evolution of $1\,-\,\mathcal{F}$ one can appreciate the power-law and linear 
time regimes, provided by the position of the interaction from dipole $l=129$ with time, given by $x(t)\,=\,129\,\pm\,40\,t^{1/3}$ and $141.5\,\pm\,57\,t$, respectively. This evolution is virtually the same 
corresponding to the configuration as depicted in Fig. 1. Positions are given in units of $a$. See text for details.}
\label{fig2}
\end{center}
\end{figure}

{\it Conclusions}.-- We have shed new light on the existence of light cones in classic linear chains of interacting dipoles. By considering two different dipole configurations, we obtain a non-linear 
light cone that goes as $\propto t^{1/3}$ for very short initial times, which gives rise afterwards to a linear one, usually for $t\,\ge\,0.1$. The conciliation of having a speed of information propagation 
less than the speed of light is possible by introducing a retarded action \`a la Li\' enard-Wiechert in the Landau-Lifshitz equation.
Regarding the connection between long ranged interactions and quantum physics,
the basic principle behind is that symmetry rules in quantum physics mean
effective long ranged interactions which are evidenced in
superconductivity, superfluidity, and so on such as the second sound in
helium, albeit there is not a direct Hamiltonian for it.
Finally, we conjecture that $A$ and $v_s$ in $A\,t^{1/3}$, $B\,+\,v_s\,t$ solely depends on universal constants, as well as the exponent 3 in $1/r^3$. This conclusion easily extends to 
similar interactions of the type $1/r^{\alpha}$.

{\it Acknowledgements}.-- J. Batle acknowledges fruitful discussions with J. Rossell\'o, Maria del Mar Batle and Regina Batle. JB received no funding for the present research.

\newpage
\vspace{5em}
\centerline{\Large\textsc{Supplementary material}}
\section{Integration method}

Let us consider the second kind of dipole configuration described in the present work. Initial conditions were set as such: a total number of 1024 dipoles are set on the $x$ axis, all of them parallel to $\mathbf{e}_x$, except for $n=512$ which is set parallel to $\mathbf{e}_z$. The propagation of this perturbation can be monitored by observing the normal component $S_\mathcal{N}=\sqrt{s_{y_n}^2(t)+s_{z_n}^2(t)}$ of the dipoles. 

\begin{figure*}[htb!]
\includegraphics[width=\textwidth]{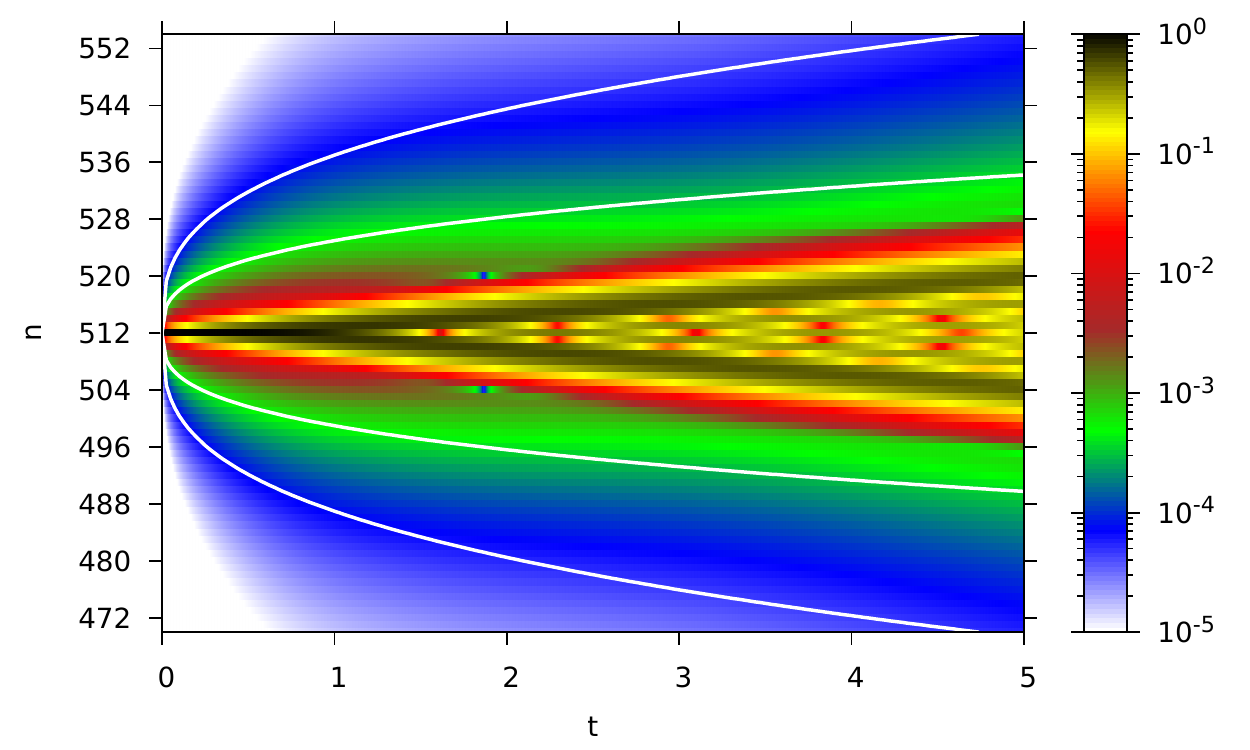}
\caption{Normal components $S_{\mathcal{N}}$ of $\mathbf{s}_n(t)$ (logarithmic color-scale) for the central dipoles at the onset of the motion. White lines indicate $t^{\frac{1}{3}}$ behavior.}\label{Syz_log_fig}
\end{figure*}
This is shown in Fig. \ref{Syz_log_fig}. A precursor is also revealed by the logarithmic color-scale. The white lines are, at this stage, guides for the eye, with a $t^{\frac{1}{3}}$ time-dependence.

The instantaneous motion generated by the Landau-Lifshitz equation is a precession motion around the field $\mathbf{H}_n$: standard integration methods, such as Euler or Runge-Kutta, do not 
usually conserve $|\mathbf{s}_n|$, so one adequate method is to use the Rodrigues rotation matrix $\mathbf{R}_n(t)$ \cite{rodrigues}, so that:
$$\mathbf{s}_n(t+ht) = \mathbf{R}_n(t)\,\mathbf{s}_n(t)$$
$ht$ being the integration time-step, with
$$\mathbf{R}_n(t) =
\left( \begin{array}{ccc}
h_x^2w+v & h_xh_yw-h_zu&h_xh_zw+h_yu\\
h_xh_yw+h_zu & h_y^2w+v  & h_yh_zw-h_xu\\
h_xh_zw-h_yu& h_yh_zw+h_xu&h_z^2w+v
\end{array}\right)$$
where $u =\sin\omega$, $v=\cos\omega$, $w=1-v$, while the precession angle is $\omega =\left|\mathbf{H}_n\right| ht$.
Coordinates $h_{x,y,z}$ are those of the unit vector $\mathbf{h}_n = \mathbf{H}_n/\left|\mathbf{H}_n\right|$. The so-called ``improved Euler'' or Heun method \cite{nowak} can then be used to integrate the Landau-Lifshitz equation.

The dipole-dipole interaction being long-ranged, the sum on $\ell$ for the field in the Landau-Lifshitz equation is bound to be costly: writing the interaction as a convolution and using fast Fourier transforms \cite{nowak} and zero-padding considerably accelerates computations.

The simulations were done without damping nor temperature regulation; the time-step $ht=2.5\,10^{-3}$ was chosen to ensure energy conservation. The use of smaller $ht$ ($10^{-5}$) is important for 
a greater time-resolution at the very beginning of the time evolution provided by the Landau-Lifshitz equation. 

\section{Precursor analysis method}

\begin{figure*}[htb!]
\includegraphics[width=\linewidth]{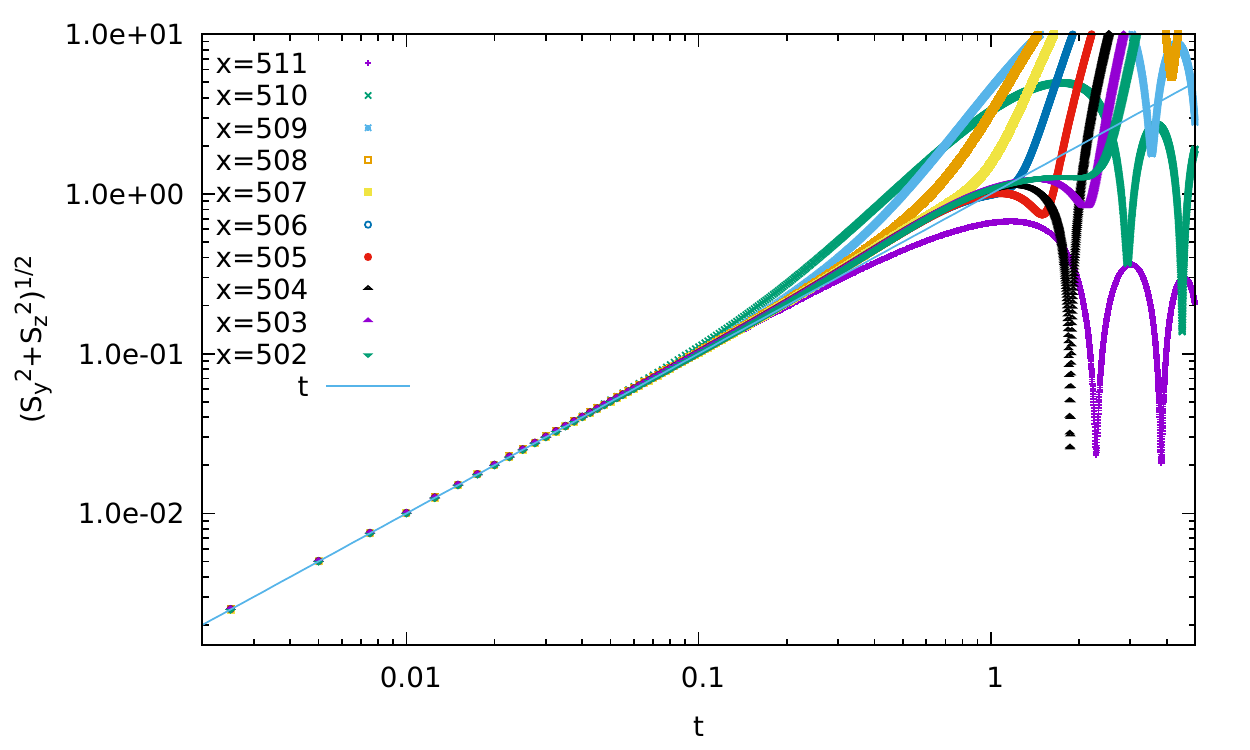}
\caption{Master plot of $S_{\mathcal{N}}$ for neighbors of the central perturbated dipole. Both scales are logarithmic. All curves are scaled to fit on the same line.} \label{Syz_master_fig}
\end{figure*}

To analyze the precursor, one cannot simply rely upon observing one given contour line. To ensure that all dipoles with measurable motion have the same behavior, a master 
plot (Fig. \ref{Syz_master_fig}) was done by rescaling $S_\mathcal{N}$ to obtain superposition. This actually occurs within the interval $[0,0.1]$ approximately. 
The master line is straight, proportional to $t$. The scaling coefficient turns out to be proportional $x^{-3}$ where $x$ is the distance from the perturbation at $n=512$. 
This can be summarized as such:
$$S_\mathcal{N}(x,t) = \frac{a t}{x^3}$$\newline
where $a$ is a constant. A given contour line is characterized by $S_\mathcal{N} = C$ a constant. Thus
$$x = \left(\frac{a t}{C}\right)^\frac{1}{3}$$
Therefore, the distance covered by a contour line is proportional to $t^\frac{1}{3}$ as could be guessed from Fig. \ref{Syz_log_fig}.

\end{document}